\documentstyle[pra,aps,multicol,psfig]{revtex}
\begin{document}
\draft
\title{Trap environment effects over the quantum statistics and the atom-photon
correlations in the Collective Atomic-Recoil Laser}
\author{G.A. Prataviera}
\address{Departamento de F\'{\i}sica, Universidade Federal de S\~{a}o Carlos. \\
Via Washington Luis Km 235, S\~{a}o Carlos, 13565-905, SP, Brazil}
\date{\today}
\maketitle

\begin{abstract}
We consider the effects of the trap environment on the atomic and optical
quantum statistical properties and on the atom-photon correlations in the
Collective Atomic-Recoil Laser. Atomic and optical statistical properties,
as well as the atom-photon correlations are dependent on the optical field
intensity and phase. In particular, depending on the values of the optical
field intensity and phase, the fields statistics varies from coherent to
superchaotic.
\end{abstract}


%\begin{multicols}{2}

%\section{Introduction}

Realization of Bose-Einstein condensation in trapped atomic gases \cite{r1}
has produced new advances in atom optics. Particularly, the interaction of
condensates with single mode quantized light fields has been a challenging
topic \cite{r7,r5,r6,r8}, allowing for instance light and matter wave
amplification \cite{r5,r6,r8} and optical control of atomic statistical
properties \cite{r5,r8}.

It is known that the trap environment can modify the properties of ultracold
atoms, such as its critical temperature \cite{c1}. In Ref. \cite{c6}, the
trap environment effect on the Condensate Collective Atomic-Recoil Laser
(CARL) \cite{c7} was included by expanding the matter-wave field in the trap
matter-wave modes. Such a situation was called Cavity Atom Optics (CAO), in
analogy with the Cavity Quantum Electrodynamics (CQE) where the spontaneous
emission is modified by the presence of the cavity. The dynamics of this
model was compared \cite{c6} with its counterpart free space model \cite{r8}
where the matter wave field was expanded in plane waves. It was found that
in the CAO regime the atomic and optical fields intensities present two
regimes of exponential instability. However, the trap effect over the
quantum statistical properties and the atom-photon correlations was not
considered. These properties are important to characterize the role of trap
environments on the control of atomic and optical statistical properties,
which can be useful for quantum information processing.

This Brief Report concerns with the effects of trap environment over the
quantum statistical properties and the atom-photon correlations in the CARL.
By analyzing the second order correlation functions and intensity cross
correlations we verify that the statistical properties of the atomic and
optical fields as well as the atom-photon correlations are dependent of the
optical field initial intensity and phase. Furthermore, by setting the
optical field intensity and phase the fields statistics can be varied from
coherent to superchaotic.

Following Ref. \cite{c6} we consider a Schr\"{o}dinger field of
non-interacting bosonic two-level atoms interacting with two single-mode
running wave optical fields of frequencies $\omega _{1}$ and $\omega _{2}$,
both being far off-resonant from any electronic transition. By two-photon
virtual transitions, in which the atoms internal state remains unchanged,
the center-of-mass motion may change due to the recoil. In the far-off
resonant regime the excited state population and spontaneous emission may be
neglected and the ground state atomic field evolves coherently under the
effective Hamiltonian 
\begin{equation}
\hat{H}=\int d^{3}{\bf r\;}\hat{\Psi}^{\dagger }({\bf r})\left[ -\frac{\hbar
^{2}}{2m}\nabla ^{2}+V({\bf r})+\hbar \left( \frac{g_{1}^{\ast }g_{2}}{%
\Delta }\hat{a}_{1}^{\dagger }a_{2}e^{-i{\bf K\cdot r}}+\frac{g_{2}^{\ast
}g_{1}}{\Delta }a_{2}^{\ast }\hat{a}_{1}e^{i{\bf K\cdot r}}\right) \right] 
\hat{\Psi}({\bf r})+\hbar (\omega _{1}-\omega _{2})\hat{a}_{1}^{\dagger }%
\hat{a}_{1}  \label{mod1}
\end{equation}
where $m$ is the atomic mass, $V(r)$ is the trap potential, $g_{1}$ and $%
g_{2}$ are the probe and pump coupling coefficients, respectively, and ${\bf %
K}={\bf k}_{1}-{\bf k}_{2}$ is the difference between the probe and pump
wavevectors. The operator $\hat{a}_{1}$ is the photon annihilation operator
for the probe mode, taken in the frame rotating at the pump frequency $%
\omega _{2}$. The pump is treated classically and assumed to remain
undepleted, and the index of refraction of the atomic sample was assumed
equal to the vacuum by neglecting spatially independent light shifts
potentials.

Now we distinguish between free propagation and cavity regimes. The free
propagation regime is valid for times short enough that atoms at the recoil
velocity propagate only over short distances compared to the dimensions of
the initial condensate. Following Ref \cite{r8}, in this regime the atomic
field is expanded onto momentum side modes, which are simply plane waves
with a slowly varying spatial envelope. The cavity regime occurs at much
longer time scales, when the recoiling atoms propagate over distances larger
than the trap dimensions. In the CAO regime the atoms probe the trap
environment and the atomic field operator is best expressed in terms of the
trap eigenmodes $\left\{ \varphi _{n}({\bf r})\right\} $ according to $\hat{%
\Psi}({\bf r})=\sum_{n=0}^{\infty }\varphi _{n}({\bf r})\,\hat{c}_{n}$,
where $\hat{c}_{n}$ is the annihilation operator for atoms in mode $n$.

We consider that the condensate mode $\hat{c}_{0}$ is initially highly
populated with a $N$ mean number of atoms, allowing treat it as a c-number.
Also is neglected condensate depletion, which is valid for short times so
that the side mode populations remain small compared to $N$. These
conditions allow the replacement $\hat{c}_{0}\approx \sqrt{N}$. For
simplicity we assume that the condensate mode couples only to the $m$th trap
mode. Thus we obtain the following effective Hamiltonian 
\begin{equation}
\hat{H}=\hbar \omega _{m}\hat{c}_{m}^{\dagger }\hat{c}_{m}+\hbar \delta \hat{%
a}^{\dagger }\hat{a}+\hbar \chi _{m}\left( \hat{a}^{\dagger }\hat{c}%
_{m}^{\dagger }+\hat{a}^{\dagger }\hat{c}_{m}+\hat{c}_{m}^{\dagger }\hat{a}+%
\hat{c}_{m}\hat{a}\right) ,  \label{lr3}
\end{equation}
where $\hbar \omega _{m}$ is the energy of the $m$th trap mode, $\delta =$ $%
\omega _{1}-\omega _{2}$ is the detuning between the pump and probe optical
fields, $\chi _{m}=\sqrt{N}A_{0m}|g_{1}||g_{2}||a_{2}|/|\Delta |$ is the
coupling constant between atomic and optical fields, $\hat{a}%
=(g_{1}g_{2}^{\ast }a_{2}^{\ast }\Delta /\left| g_{1}\right| \left|
g_{2}\right| \left| a_{2}\right| \left| \Delta \right| )\hat{a}_{1}$ is the
probe annihilation operator times a phase factor related to the phase of the
pump laser and the sign of the detuning, and $A_{0m}=\int d^{3}{\bf r}%
\varphi _{0}^{\ast }({\bf r})e^{i{\bf K\cdot r}}\varphi _{m}({\bf r})$ is
the matrix element for the optical transition that was assumed a real
number. Terms like $\hat{a}^{\dagger }\hat{c}_{m}^{\dagger }$ in (\ref{lr3})
correspond to the generation of correlated atom-photon pairs. This is
analogous to the Optical Parametric Amplifier (OPA), except that in that
case it is correlated photon pairs that are generated.

The Heisenberg equations of motion for the field operators follow from
Hamiltonian (\ref{lr3}), resulting the $4\times 4$ system of equations 
\begin{equation}
\frac{d}{dt} \left( 
\begin{array}{c}
\hat{c}_{m} \\ 
\hat{c}_{m}^{\dagger } \\ 
\hat{a} \\ 
\hat{a}^{\dagger }
\end{array}
\right) =i\left( 
\begin{array}{cccc}
-1 & 0 & -\chi _{m} & -\chi _{m} \\ 
0 & 1 & \chi _{m} & \chi _{m} \\ 
-\chi _{m} & -\chi _{m} & -\delta & 0 \\ 
\chi _{m} & \chi _{m} & 0 & \delta
\end{array}
\right) \left( 
\begin{array}{c}
\hat{c}_{m} \\ 
\hat{c}_{m}^{\dagger } \\ 
\hat{a} \\ 
\hat{a}^{\dagger }
\end{array}
\right)  \label{lr2}
\end{equation}
where we introduced the dimensionless quantities $t\equiv\omega_{m}t $, $%
\delta \equiv\delta /\omega _{m}$ and $\chi _{m}\equiv\chi _{m}/\omega _{m}$.

The solution of the linear system (\ref{lr2}) can be written as 
\begin{equation}
\hat{x}_{i}(t)=\sum_{j=1}^{4}\sum_{k=1}^{4}G_{ij}^{(k)}(t)e^{i{\bf \omega }%
_{k}t}\,\hat{x}_{j}(0),  \label{sta1}
\end{equation}
where we defined $\hat{x}_{1}=\hat{c}_{m}$, $\hat{x}_{2}=\hat{c}%
_{m}^{\dagger }$, $\hat{x}_{3}=\hat{a}$ and $\hat{x}_{4}=a^{\dagger }$ for
convenience, and $\omega _{k}$ are the system eigenfrequencies. For
non-degenerate eigenfrequencies $G_{ij}^{(k)}(t)=\left[ {\bf U}\right] _{ik}[%
{\bf U}^{-1}]_{kj}$, where $\left[ {\bf U}\right] _{ik}$ is \ the $i$th
component of the $k$th eigenvector of the matrix at RHS of Eq. (\ref{lr2}).
Following \cite{c6}, the stability analysis shows three distinct regimes:
(i) The eigenfrequencies are purely real, the system being stable. (ii) Two
purely real and two purely imaginary eigenfrequencies of the form $\left\{
\omega _{1}=\Omega ,\omega _{2}=-\Omega ,\omega _{3}=i\Gamma ,\omega
_{4}=-i\Gamma \right\} $, where $\Omega $ and $\Gamma $ are both real
quantities. There is only one exponentially growing solution at the
imaginary frequency $\omega _{4}$ and the system is unstable. (iii) The
eigenfrequencies are complex numbers of the form $\left\{ \omega _{1}=\Omega
+i\Gamma ,\omega _{2}=-\Omega +i\Gamma ,\omega _{3}=\Omega -i\Gamma ,\omega
_{4}=-\Omega -i\Gamma \right\} $. This case presents two exponentially
growing solutions, $\omega _{3}$ and $\omega _{4}$, which grow at the same
rate $\Gamma $, but rotate at equal and opposite frequencies $\pm \Omega $,
producing a beating in the exponential growth of the fields intensities. A
fourth possibility is: (iv) The eingenfrequencies are not imaginary and are
degenerate at the critical values $\delta =0$, $\delta =4\chi _{m}^{2}$ and $%
\left( 1-\delta ^{2}\right) ^{2}/\left| \delta \right| =16\chi _{m}^{2}$ ($%
\delta <0$), which define the threshold separating stable from exponentially
unstable solutions. The coefficient $G_{ij}^{(k)}(t)$ is a polynomial in $t$
of degree one less than the degree of degeneracy. The fields amplitudes
acquire a linear time dependence and the system becomes unstable. We are
interested in the unstable regimes (ii), (iii) and (iv), which show
stationary statistical characteristics as we see below.

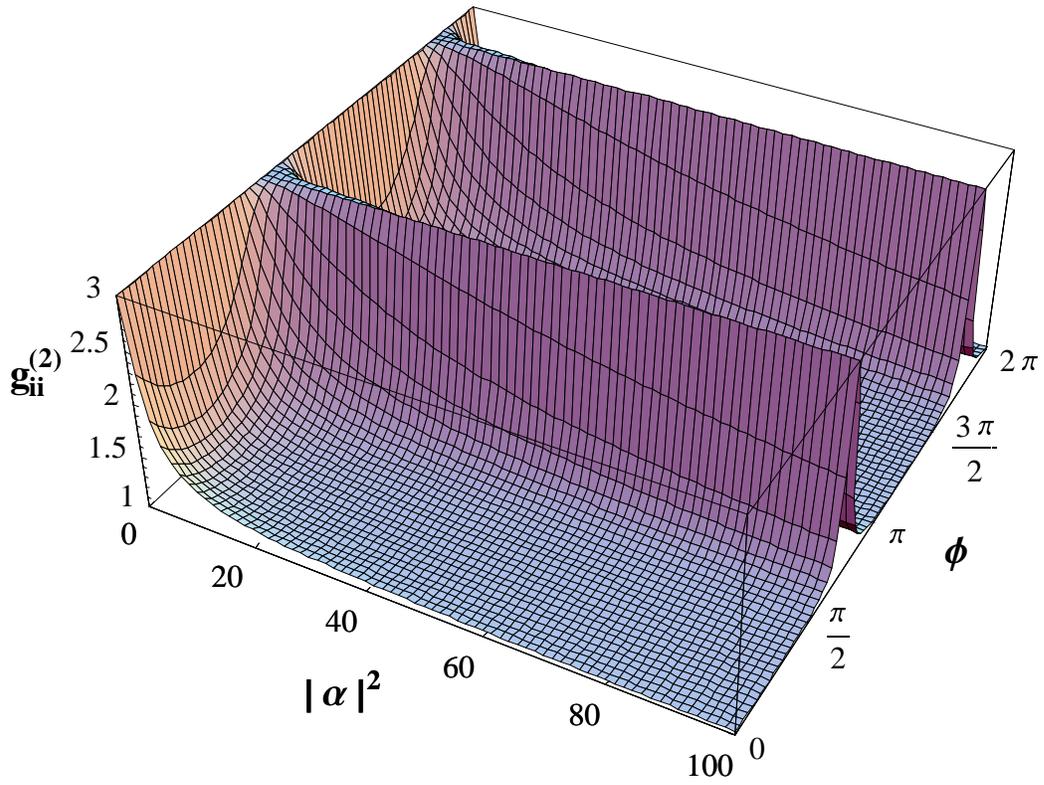
\begin{figure}[tbp]
\caption{Plot of the long time value of atomic and optical fields second
order correlation function as a function of $|\protect\alpha |^2$ and $%
\protect\phi$ in the regime (ii). The parameters are set $\protect\delta=1 $
and $\protect\chi_m=1$. }
\label{f1}
\end{figure}

Let us first consider the atomic and optical fields statistical properties,
which can be characterized by the normalized equal-time second-order
correlation function defined by \cite{e1} 
\begin{equation}
g_{ii}^{(2)}(t)=\frac{\left\langle \hat{x}_{i}^{\dagger }(t)\hat{x}%
_{i}^{\dagger }(t)\hat{x}_{i}(t)\hat{x}_{i}(t)\right\rangle }{\left\langle 
\hat{x}_{i}^{\dagger }(t)\hat{x}_{i}(t)\right\rangle ^{2}},\quad i=1,3.
\label{st1}
\end{equation}
We assume that the atomic mode begin in a vacuum state whereas the light
field is initially in a coherent state $|\alpha \rangle $ with complex
amplitude $\alpha =\left| \alpha \right| e^{-i\phi }$. The calculated
expression $g_{ii}^{(2)}(t)$ is dependent on the optical field intensity $%
\left| \alpha \right| ^{2}$ and phase $\phi $, as well as of the system
parameters. Unfortunately, the analytical expression does not provide so
much insight. We begin by considering values of parameters in the regime
(ii). After a transient, the exponential with eigenfrequencie $\omega _{4}$
will dominate in the Eq. \ref{sta1} and both, $g_{11}^{(2)}(t)$ and $%
g_{33}^{(2)}(t)$, attains at long time the same constant value. The long
time value of $g_{ii}^{(2)}(t)$ is plotted in Fig. \ref{f1} as function of $%
\left| \alpha \right| ^{2}$ and $\phi $, which shows a strong sensitivity to
the initial phase and intensity of the light field. We see that by varying
the light field intensity and phase it is possible to continuously change
the fields statistics from coherent ($g_{ii}^{(2)}=1$), to chaotic ($%
g_{ii}^{(2)}=2$), to superchaotic ($g_{ii}^{(2)}>2$). The generation of
superchaotic light was first analyzed in Ref. \cite{w1} by considering a
two-photon emission process. The situation considered here is rather
interesting because it allows to produce a superchaotic atomic source.

In the regime (iii) the atomic and optical correlation functions attain at
long times a stationary oscillating values, which are not necessarily the
same. In order to explore the long time optical field intensity and phase
sensitivity it is plotted in Fig. \ref{f2}-(a) and \ref{f2}-(b) the value of 
$g_{11}^{(2)}$ and $g_{33}^{(2)}$, respectively, as a function of $\left|
\alpha \right| ^{2}$ and $\phi $ by considering a fixed large value of time.
We see, by comparing Fig. \ref{f1} with Figs. \ref{f2} that the later one
shows less sensitivity to the optical field phase. In the regime (iv) the
correlations attains at long time a steady value with amplitude decreasing
oscillations around a constant value for $\delta =0$ and $\delta =4\chi
_{m}^{2}$, and oscillating for $\left( 1-\delta ^{2}\right) ^{2}/\left|
\delta \right| =16\chi _{m}^{2}$ ($\delta <0$). For $\delta =0$ and $\delta
=4\chi _{m}^{2}$ an analytical expression for the asymptotic value of $%
g_{ii}^{(2)}(t)$ can be found by maintaining only those terms with linear
time dependence in the fields amplitudes. Both $g_{11}^{(2)}$ and $%
g_{33}^{(2)}$ attains the same value given by 
\begin{equation}
g_{ii}^{(2)}=1+2\frac{\left[ 1+\delta _{c}\right] \left[ 1+\delta
_{c}+8\left| \alpha \right| ^{2}\cos ^{2}(\phi -\frac{\pi \delta _{c}}{8\chi
_{m}^{2}})\right] }{\left[ 1+\delta _{c}+4\left| \alpha \right| ^{2}\cos
^{2}(\phi -\frac{\pi \delta _{c}}{8\chi _{m}^{2}})\right] ^{2}},  \label{st4}
\end{equation}
where $\delta _{c}=0$ or $\delta _{c}=4\chi _{m}^{2}$. From Eq. (\ref{st4})
we see that $1\leq g_{ii}^{(2)}(t)\leq 3$, which confirm the presence of
superchaotic fluctuations. The behavior of $g_{ii}^{(2)}(t)$ at $\left(
1-\delta ^{2}\right) ^{2}/\left| \delta \right| =16\chi _{m}^{2}$ ($\delta
<0 $) is similar to that found in the regime (iii), however no simple
analytical expression was obtained.

Now we turn to analyze the atom-photon correlations, which be quantified by
the two-mode equal-time intensity cross-correlation function \cite{e1} 
\begin{equation}
g_{ij}^{(2)}(t)=\frac{\left\langle \hat{x}_{i}^{\dagger }(t)\hat{x}_{i}(t) 
\hat{x}_{j}^{\dagger }(t)\hat{x}_{j}(t)\right\rangle }{\left\langle \hat{x}
_{i}^{\dagger }(t)\hat{x}_{i}(t)\right\rangle \left\langle \hat{x}
_{j}^{\dagger }(t)\hat{x}_{j}(t)\right\rangle },\quad i\neq j.
\label{crossc1}
\end{equation}
In particular, for classical fields the two-mode correlation function is
bounded by 
\begin{equation}
g_{ij}^{(2)}(t)\leq \sqrt{g_{ii}^{(2)}(t)g_{jj}^{(2)}(t)},  \label{crossc2}
\end{equation}
while quantum fields can violate this inequality, being limited by 
\begin{equation}
g_{ij}^{(2)}(t)\leq \sqrt{\left[ g_{ii}^{(2)}(t)+\frac{1}{\left\langle \hat{x%
}_{i}^{\dagger }(t)\hat{x}_{i}(t)\right\rangle }\right] \left[
g_{jj}^{(2)}(t)+\frac{1}{\left\langle \hat{x}_{j}^{\dagger }(t)\hat{x}%
_{j}(t)\right\rangle }\right] },  \label{crossc3}
\end{equation}
which reduces to the classical limit at large intensities.

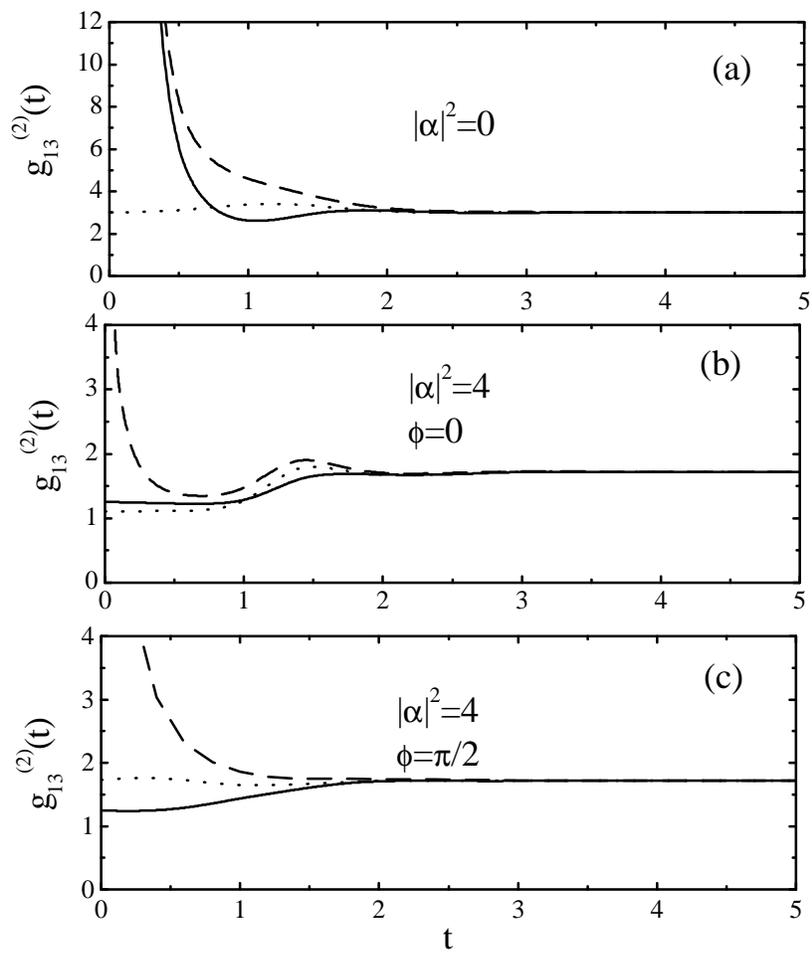
\begin{figure}[tbp]
\caption{Solid line: plot of the atom-photon second order cross-correlation
function as a function of time in the regime (ii). Dashed line: quantal
upper limit. Doted line: classical upper limit. The parameters are set $%
\protect\delta =1$ and $\protect\chi _{m}=1$.}
\label{f3}
\end{figure}

In order to explore the general characteristics of the atom-photon
correlations we consider only the exponentially ustable regimes (ii) and
(iii). Considering the regime (ii), the sequence in Fig. \ref{f3} shows the
atom-photon correlation function \ $g_{13}^{(2)}(t)$ as a function of time
and by considering different values of light field intensity and phase. For
comparison, in the same figure it is also plotted the classical upper limit
given by Eq.\ref{crossc2} (dotted line) and the quantum upper limit given by
Eq. \ref{crossc3} (dashed line). Fig. \ref{f3}(a) shows the spontaneous case
($\alpha =0$), in which the fields instability are triggered by the noise
from vacuum fluctuations. We see that only at short times the violation of
the classical inequality is close to the maximum limit consistent with
quantum mechanics. In the case that $\alpha \neq 0$ the correlations are
also phase-dependent. Figs. \ref{f3} (b) and \ref{f3}(c) show a situation
with same intensity $|\alpha |^{2}=4$ , but taking two different values of
phase. We see that even at short times, a non-zero intensity reduces the
correlations for values close to the classical ones. The phase dependence is
evident by comparing Figs. \ref{f3}(b) and \ref{f3}(c), where the later one
presents no violation of the classical inequality. At long times the
correlations attains the classical limit and the fields become uncorrelated (%
$g_{ij}^{(2)}=1$) or correlated ($g_{ij}^{(2)}>1$) depending on the optical
field intensity and phase.

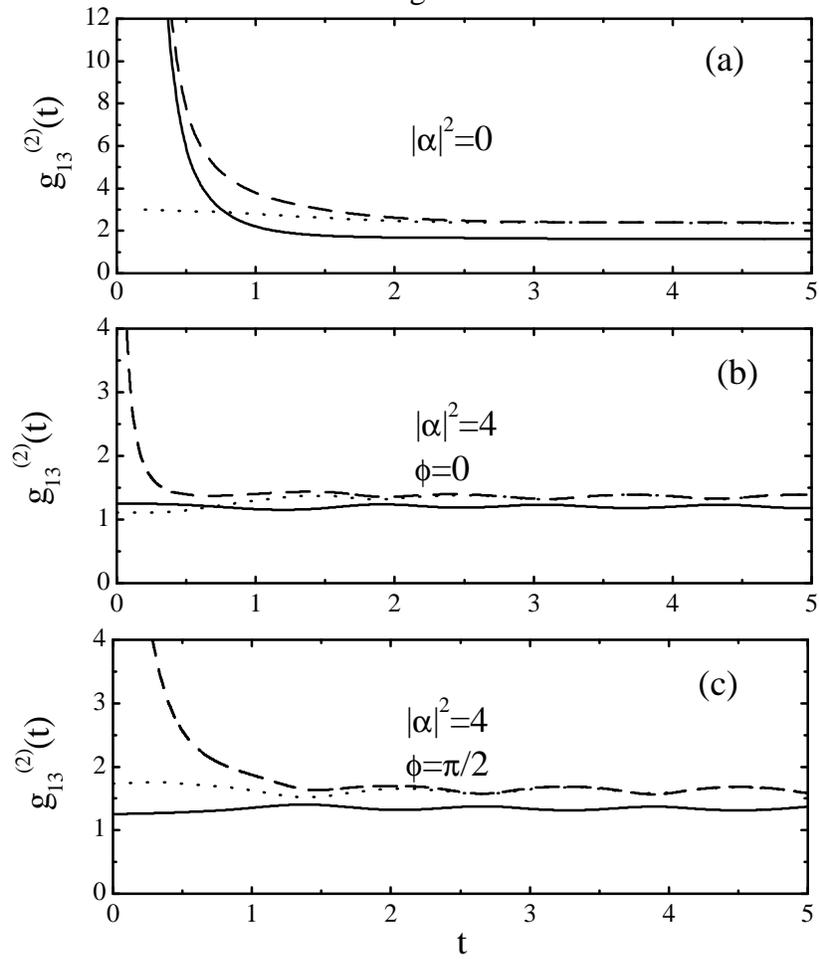
\begin{figure}[tbp]
\caption{Same as in Fig. \ref{f3} in the regime (iii). The parameters are
set $\protect\delta =-1$ and $\protect\chi _{m}=1$.}
\label{f4}
\end{figure}

The regime (iii) presents, at short and intermediate times, similar
characteristics to the regime (ii). The main difference is the long time
limit where the correlations attain a stationary value below the classical
upper limit. Figs. \ref{f4}-(a)-(b)-(c) illustrate the behavior of $%
g_{13}^{(2)}(t)$ as a function of time and its dependence to the optical
field intensity and phase.

Note that in the two-mode OPA the difference of population between the two
modes is a constant of motion \cite{e1}, revealing that in the spontaneous
case ($\alpha =0$) the two-mode correlation function shows the maximum
violation of the classical inequality consistent with quantum mechanics.
There is no such a constant of motion in the model considered in this paper
because both the process of emission and absorption of photons transfer
atoms to the same quantum state. Thus justifying that the violation of the
classical inequality in Figs. \ref{f3}(a) and \ref{f4}(a) are not maximized
for all times. Furthermore, although the violation of classical inequality
indicates nonclassicality, it is not possible to determine whether this
represents an entanglement of the two modes or if each mode is in a
nonclassical state. Further indication of entanglement is given by the
presence of chaotic intensity fluctuations in the individual modes, which
means that each of its components is in a mixed state, although the
atom-photon system is in a pure state.

In summary, we studied the effects of the trap environment on the quantum
statistical properties and on the atom-photon correlations in the CARL. The
atomic and optical fields statistical properties and the atom-photon
correlations are sensitive to the initial intensity and phase of the light
field. This result is contrasting when compared with the counterpart free
space regime model \cite{r8} whose statistical properties are only dependent
on the intensity of the light field. Furthermore, from the viewpoint of
optical control of statistical properties of atoms, while in the free space
regime the variation is limited between coherent and chaotic, in the CAO
regime also it is possible to reach a superchaotic statistics by setting the
optical field intensity and phase.

\vspace{0.1cm} %\acknowledgments{
Thanks to V. V. Dodonov and M. C. de Oliveira for useful discussions and
FAPESP-Brazil for financial support. %}

%\bigskip

%\end{multicols}

\vspace{-0.5cm}

\vspace{-0.5cm}

\begin{figure}[tbp]
\caption{Same as in Fig. \ref{f1} in the regime (iii) and setting $t=8$. (a)
atomic field, (b) optical field. The parameters are set $\protect\delta=-1$
and $\protect\chi_m=1$.}
\label{f2}
\end{figure}

\vspace{-0.5cm}

\vspace{-0.5cm}

\bigskip

\bigskip

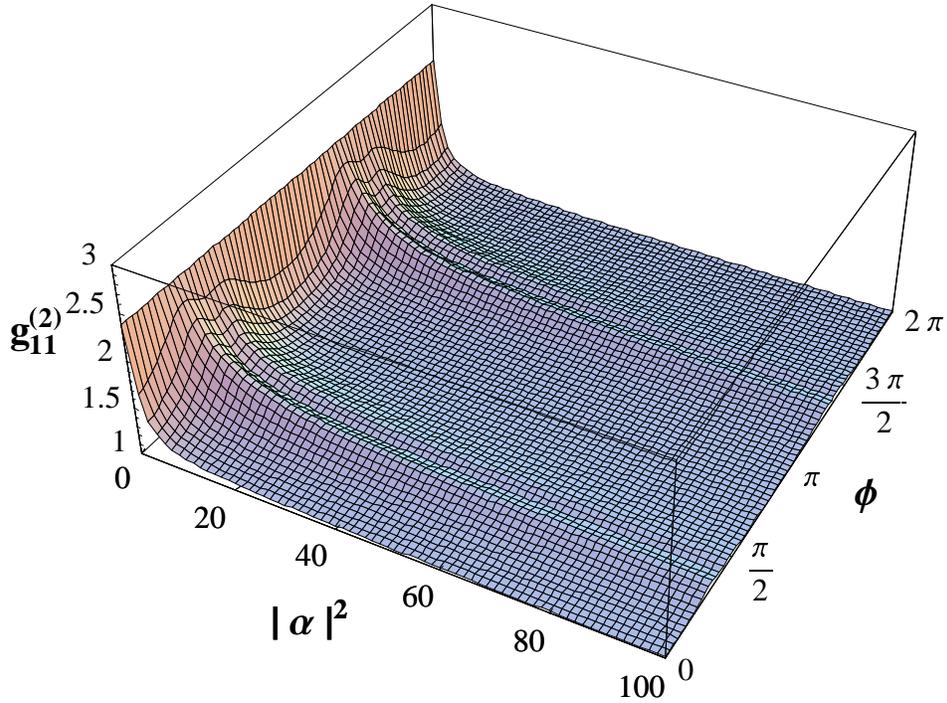
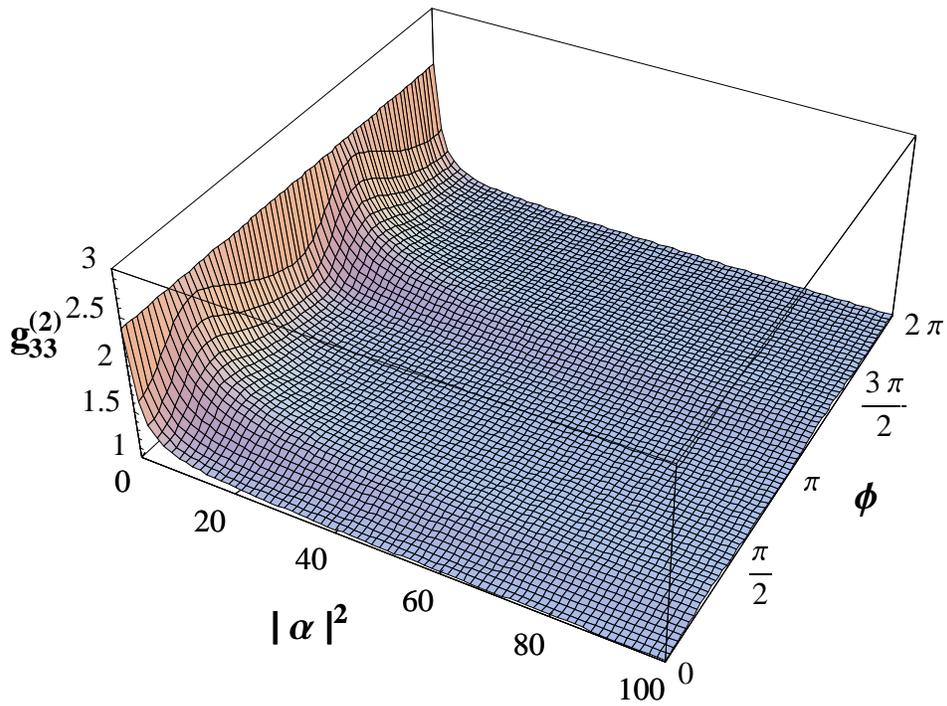
\end{document}